%% file: paper.tex
\documentclass[copyright,creativecommons]{eptcs}

\usepackage{iftex}

\ifpdf
  \usepackage{underscore}         
  \usepackage[T1]{fontenc}        
  \usepackage{xcolor}
\else
  \usepackage{breakurl}           
\fi

\usepackage[nameinlink,capitalise]{cleveref}

\title{Towards Richer Challenge Problems for Scientific Computing Correctness}
\author{Matthew Sottile
\and
Mohit Tekriwal
\institute{Lawrence Livermore National Laboratory, USA}
\email{\{sottile2, tekriwal1, sarracino2\}@llnl.gov}
\and
John Sarracino
}

\newcommand{\titlerunning}{Towards Richer Challenge Problems for Scientific Computing Correctness}
\newcommand{\authorrunning}{M. Sottile, M. Tekriwal \& J. Sarracino}

\hypersetup{
  bookmarksnumbered,
  colorlinks,
  linkcolor=red!50!black,
  citecolor=blue!50!black,
  urlcolor=blue!80!black,
  pdftitle    = {\titlerunning},
  pdfauthor   = {\authorrunning},
  pdfsubject  = {EPTCS},               
}

\begin{document}
\maketitle

\begin{abstract}
Correctness in scientific computing (SC) is gaining increasing
attention in the formal methods (FM) and programming languages (PL)
community. Existing PL/FM verification techniques struggle with the
complexities of realistic SC applications. Part of the problem is a
lack of a common understanding between the SC and PL/FM communities of
machine-verifiable correctness challenges and dimensions of
correctness in SC applications.

To address this gap, we call for specialized challenge problems to
inform the development and evaluation of FM/PL verification techniques
for correctness in SC. These specialized challenges are intended to
augment existing problems studied by FM/PL researchers for general
programs to ensure the needs of SC applications can be met. We propose
several dimensions of correctness relevant to scientific computing,
and discuss some guidelines and criteria for designing challenge
problems to evaluate correctness in scientific computing.
\end{abstract}


\input{benchmark}

\section*{Acknowledgements}
This manuscript has been authored by Lawrence Livermore National Security, LLC under Contract No.
DE-AC52-07NA2 7344 with the US. Department of Energy. The United States Government retains, and
the publisher, by accepting the article for publication, acknowledges that the United States Government
retains a non-exclusive, paid-up, irrevocable, world-wide license to publish or reproduce the published
form of this manuscript, or allow others to do so, for United States Government purposes.

\bibliographystyle{eptcs}
\bibliography{references}
\end{document}

%% file: benchmark.tex

\section{Introduction}

Scientific computing and high performance
computing has long relied upon benchmark suites to give computer
science researchers target challenge problems that are known to be
relevant to end-users in the sciences. These have most often been
successful in the area of performance benchmarks that have driven
compiler, systems, and parallel computing research. For example, the
NAS Parallel Benchmarks (NPB)~\cite{npb1,npb2} were created to be
representative of highly parallel HPC problems that arise in aerospace
applications, specifically in fluid dynamics simulations. The Mantevo
problems~\cite{mantevo} provide similar problem collection
representative of those of interest to the broad HPC community within
the US Department of Energy. More specific suites aimed
at compiler and language researchers include the Salishan
problems~\cite{salishan} and the more recent Shonan
challenge~\cite{shonan}.

The challenge problems shared by the HPC community have grown
in sophistication over time tracking with two key aspects of the 
research field.  First, challenges grew more complex as application
codes and computers to run them grew more complex.  Second, computer
scientists studying optimization techniques and architectures were able to
effectively handle simple benchmarks and needed more complexity to push
their own research forward.  This evolution of complexity can be seen in
the steps from the earliest linear algebra benchmarks like
Linpack~\cite{linpack} through more specialized loop kernels in the
Livermore Loops~\cite{mcmahon1986livermore}, eventually leading to algorithmic and application
kernels that started simple (NAS Parallel Benchmarks) but grew to current
suites that contain nearly complete application codes or mini-apps (Mantevo).

The constraints that accompanied these changed as well: early challenges focused on
simple metrics like floating point operations per second, and eventually integrated
correctness metrics (especially relevant for aggressive optimization techniques) and
well defined methods for scaling problems (e.g., the NAS Parallel Benchmark problem class sizes).
These additions were necessary to give optimization and performance researchers tools to
evaluate their work more strongly than simply looking at raw performance numbers. We believe that the 
correctness challenge problems should take inspiration from this
evolution of performance challenges to drive the field of software correctness forward
to ever more complex and realistic problem areas with a rich set of correctness criteria
to reason about.

A key focus when constructing the performance-oriented benchmark
suites was to adequately cover the set of known computer and software
design dimensions that a system exhibits that influence performance.
These range from memory traversal patterns and the relationship to the
memory hierarchy of the computer, floating point load, fine grained instruction
level parallelism, and so on.   In the performance of scientific computing applications, these
dimensions that describe different applications are relatively well
understood and have motivated most of the benchmark suites we see
today. We believe that this remains an open question in the correctness context: 
\emph{what are the dimensions of correctness relevant to scientific computing 
practitioners and what suite of challenge problems will sufficiently cover them to
support researchers studying scientific computing correctness problems}?

Existing challenge problems in the formal verification community are a
good starting point. For example, many of the programs in the Gallery
of Verified Programs~\cite{gallery,inriaGalleryVerified} are drawn
from challenges like VerifyThis~\cite{verifythis}, the NSV-3
benchmarks~\cite{nsv3,nsv3journal}, and various verification
competitions. 
These have gaps though: classes of algorithms (e.g., graph methods that
aren't simple traversals), sophisticated data representations (e.g., sparse
matrix representations beyond basic CSR), and fundamental areas of 
mathematics all are poorly covered in current challenge and demonstration
suites.

While the numerical verification community has developed
verified numerical algorithms for decades (such as coq-num-analysis~\cite{coqnum} and the Mathematical Components Library~\cite{mathcomp}),
these individual research efforts have not yet been condensed into a general set of challenges relevant to the scientific computing community.
Building from
the existing formal verification challenges that set a good foundation would be extremely
useful to pull formal verification techniques closer to the kinds of algorithms and
programs that one encounters in practice in modern scientific
computing.

\subsection{Challenge problems vs benchmarks}

There is a distinction between
challenge problems and benchmarks. Challenge problems define a target
to reach for and a criterion to assess whether that target has been reached.
Benchmarks move a step further to not only assess that the target has been
reached, but provide an objective set of metrics that assess the relative quality between different solutions.
Current scientific computing benchmark suites have relatively narrow scopes of correctness:
floating point benchmark suites~\cite{fpbench} focus on numerical
issues, while suites such as
DataRaceBench~\cite{liao2017dataracebench} focus on specific classes
of concurrency control problems. Instead, most suites are designed to play the role of
benchmarks focusing on the metrics to achieve.  Ideally we will achieve challenge problems that
achieve both goals of providing rich correctness criteria as well as well defined metrics, but in the meantime
we believe a focus of the correctness community should emphasize defining challenges with rich correctness criteria.

\subsection{Contributions}
In this position paper,
we advocate for a set of \emph{correctness} challenge problems for scientific computing.
We develop two distinct categories of insights: 1) a set of \emph{Dimensions of Correctness} that correctness challenge problems should encompass in \cref{sec:dimensions}; and 2) a set of \emph{Pitfalls} that correctness challenge problems should \emph{Avoid} to ensure interest and relevancy to scientific computing practitioners in \cref{sec:pitfalls}.
Our recommendations are meant as a starting point to a broader discussion between the scientific computing and formal methods communities.

\section{Dimensions of Correctness}
\label{sec:dimensions}
We propose correctness dimensions that are analogous to the performance dimensions traditionally used to
select components of benchmark suites.
While some individual aspects of scientific computing correctness are well-studied by the verification community,
other aspects are not, 
and they have not been combined into a representative benchmark problem (or set of benchmark problems).
To inform the design of comprehensive correctness challenge problems,
we propose dimensions that encompass a realistic scientific application.

\paragraph{Comprehensive Dimensions}
Scientific programs straddle different layers of the computer abstraction hierarchy, 
and so the dimensions for a realistic application must encompass the relevant layers.
At the lowest level of an application we have instances of numerical calculations and traditional data structures (such as tensors),
while at the intermediate level there are model-specific data structures and computations (such as meshes or iterative solvers).
These model-specific abstractions in turn connect to high-level properties, 
such as mathematical abstractions and invariants for physical systems, and
solution methods for governing equations.

The low-level dimensions, such as numerical operations, memory management, and concurrency~/~parallelism, are broadly important and so are generally well-studied by the formal methods field.
Some existing general-purpose benchmarks in these areas include FPbench~\cite{fpbench}, DataRaceBench~\cite{liao2017dataracebench}, and SPEC~\cite{dixit1991spec}.
Notice though that such general-purpose benchmarks do not extend to higher-level layers of a scientific computing application.
The most impact on scientific computing will be to fill the
gaps between those low level aspects of a system and the high-level
applications that are built atop them.
It is also necessary to combine the (currently well-studied) low-level dimensions
with new intermediate- and high-level dimensions into standalone, comprehensive applications.
We propose the following metrics to capture correctness criteria,
which includes low-level, intermediate, and high-level application correctness properties
necessary for a compelling scientific application.

\paragraph{Numerics} Scientific computing is fundamentally about
numerical calculations and correctness often is defined as a correct
correspondence between mathematical operations and their
implementation in hardware or software. This is most apparent when
considering floating point operations that are key to nearly all
scientific applications. An additional challenge is the advent of
\emph{reduced precision} floating-point formats. With increasing
hardware support for reduced precision, realistic benchmarks must
necessarily incorporate reduced precision and mixed precision
algorithms.

\paragraph{Data structures}  Scientific codes typically must manage
a great deal of data in support of the calculations they perform.
While many standard data structures are heavily used (e.g., containers
from the C++ Standard Template Library), it is not uncommon for
applications to implement modified or hybrid data structures to allow
for finer grained performance tuning. For example, a program may
``transpose'' a data structure (such as a struct-of-arrays (SOA) to
array-of-structs (AOS)) during different phases of the code to
optimize for different memory traversal patterns. Correctness in that
case not only means that each instance of the structure (SOA, AOS) is
correct, but the two structures are semantically identical modulo
physical layout and performance changes. Similarly, priority queues,
heaps, and other structures also form key parts of codes such as
discrete event simulations. These data structures typically have well
understood correctness properties (e.g., invariants, memory safety,
etc.).

\paragraph{Domain-modeling structures}  A specialized class of data
structure that is worth calling out independently are those used for
domain modeling. Often these resemble generic data structures, but
carry with them additional constraints on correctness that derive from
the physical system that they are modeling. These range from grids and
meshes that represent a discretized view of the subject of the model
(e.g., an airplane wing), to data structures that support efficient
calculations (e.g., a space partitioning data structure used for
collision detection). These data structures range from the simple
(regular $n$-dimensional arrays) to very complex (unstructured grids,
graphs, trees, and sometimes a hybrid of all three). Correctness of
these data structures not only include generic properties (e.g.,
memory safety, invariants, structural assertions) but also must
include high-level application domain invariants (e.g., conservation
laws).

\paragraph{Differential equations}
Scientific computing applications mostly concern with modeling physics using
partial differential equations (PDEs) and then solving those PDEs approximately.
For a numerical solver to
emit a correct physical solution, consistency of the PDEs with respect to the physics being
modeled must be formalized. 
Thus, a realistic benchmark should contain
consistent formal approaches or abstractions to model these PDEs and associated correctness criteria such as
well-posedness, boundary condition compatibility, numerical stability, etc.
Such correctness metrics are well-studied in the numerical methods communities
and must necessarily be part of the correctness of a scientific computing challenge problem.

\paragraph{Concurrency and parallelism} Scientific computing is rarely a
single CPU, single computer activity - it is the home of some of the
most complex parallel programs around. Reasoning about correctness of
parallel applications is challenging because often scientific
applications combine different parallel programming and concurrency
control models in a single application. Reasoning about correctness
requires considering a combination of shared memory, vector, and
distributed memory parallelism: often implemented by different
libraries that are not necessarily aware of each other. Even worse, in
high performance scientific computing we are not just concerned with
correctness, but with performance - so it is necessary to reason about
potentially complex algorithms that attempt to minimize
synchronization in the interest of performance (at the cost of
analytical complexity).

\paragraph{Approximation schemes} Scientific programs implement
models of phenomena of interest that often rely upon some
approximation methods. These are chosen for a variety of reasons:
incomplete understanding of the system of interest, infeasibility of
running a full fidelity model, or limited availability of necessary
computational resources. Challenge problems that allow us to reason
about approximations are important to allow people to reason about the
validity or quality of approximation methods. It is important to
consider approximation schemes separately from numerical methods,
which are a very specific class of approximation that is driven by our
implementation of numbers in computers. Examples of approximation
schemes that are separate from numerical methods include algorithmic
heuristics where numerical schemes break down (e.g., around discontinuities)
or interpolation methods that provide values for parameters where no closed
form model for them exists.

\section{Pitfalls to Avoid}
\label{sec:pitfalls}
~
We next focus on the gap between traditional software verification correctness challenge problems and their relevancy to realistic scientific computing applications.
When considering performance benchmarks, one goal often was diversity
of implementation: a good benchmark suite allowed for implementation
in multiple languages and environments to ensure that they were
representative of the community. A benchmark tuned to only be suitable
for a single system or language often is less useful to the community.
For example, a challenge problem specifically designed to test the
ability of a functional language compiler to desugar a high-level
syntax to a lower level form is not useful to a language developer
working with a classical imperative language - thus it is not a
generalizable challenge problem.

\paragraph{Avoid Over-specialization} In the correctness domain it is useful to ensure that correctness
challenges also avoid over-specialization to specific techniques or
systems used to reason about correctness. A problem that is highly
specialized to SMT solvers or dependently typed proof assistants
is better suited for those communities than it is to the
broader scientific computing community. Such problems are more
suitable to challenges such as those studied as part of the SAT
competition~\cite{sat,simon2005sat2002} or static analysis test suites such as
Juliet~\cite{nist-juliet}. We should instead focus on creating
challenge problems that are sufficiently high level that they can be
mapped onto different methods for reasoning about them. It may be the
case that some problems map poorly onto some methods, but that is
perfectly reasonable for challenge problems as it provides something
for practitioners to aim at.

\paragraph{Avoid ``Toy'' Problems} It is also important to consider benchmarks that are realistic. Many
running examples of ``scientific applications'' are not well aligned with the
current state of the art. Either they represent extremely dated
methods that are now considered defunct,
originate from introductory textbooks with the intention of teaching
foundations, or they come from contexts
outside the practice of domain sciences where efficiency, accuracy,
and precision are not critical. For example, a common running example is a
direct $O(n^2)$ solver for an n-body model, but in practice more
efficient methods based on space partitioning trees (e.g., the
Barnes-Hut method~\cite{bh}) or multipole methods~\cite{fmm} are used.
These methods bear little similarity to the simpler direct method.

\paragraph{Focus on correctness issues unique to scientific applications}
Finally, good challenge problems for scientific computing focus on the issues
unique to scientific computing. While broad application of challenge problems
is useful, general purpose challenges often fail to focus on the
correctness dimensions that address issues unique or of unusually high
importance to scientific computing compared to other domains. For
example, many domains outside of SC and HPC require bit-level correctness
(e.g., cryptographic algorithms), or a very relaxed level of
correctness where differences are imperceptible (e.g., visual
effects). Scientific computing brings in richer requirements that
often fall somewhere in between: preservation of a conservation law or
non-violation of a bound on the rate by which information can
propagate (e.g, the CFL condition). Integrating these kinds of
physical correctness criteria into a benchmark suite will be very valuable in
connecting what the benchmark evaluates to the constraints that are
relevant to scientific applications.  These criteria also may inform the kinds
of techniques that can be applied, especially in the case where meeting the criteria
is dependent on runtime properties such that static methods applied to code
are insufficient.

\paragraph{Consider uncertainty}  Scientific problems have inherent
uncertainties that arise from unavoidable measurement error during experimental data collection or
the incompleteness of scientific theories being studied.  In traditional program
correctness reasoning we do not consider uncertainty in inputs and outputs, but in
the sciences this is inescapable.  Furthermore, we typically do not reason about
how uncertainty of inputs is amplified or diminished as computations proceed.
Defining correctness challenges that integrate knowledge and techniques from
the uncertainty quantification and statistical modeling community would be
beneficial to the correctness community.

\paragraph{Separate underlying mathematics from the implementation}
The fundamental underlying mathematics for various numerical
approximation schemes remains the same irrespective of their
implementation. For instance, the Gram-Schmidt algorithm fundamentally
computes an orthogonal set of vectors given a linearly independent set
of input vectors. While multiple variants exist to compute this, they all
need to satisfy the same mathematical
invariants or fundamental characteristics of the Gram-Schmidt method.
As scientific libraries
are ported and rewritten to cater to the performance benefits of
evolving hardware, we need to be able to verify these implementations
with respect to the high-level mathematical methods they are instances of. 

\paragraph{Allow unchecked assumptions}  Often programs state a set of assumptions they make about
their inputs that are necessary for the program to function correctly.
For example, a function may assume all inputs are positive, or a
simulation may assume that an input triangular or tetrahedral mesh
avoids angles below a certain threshold. Given that scientific
computing programs often emphasize performance, it is not uncommon for
code to make these assumptions but not waste compute time checking
them. We may discover that challenge problems for scientific problems
will carry more unchecked assumptions about their inputs than we may
find in other domains simply because these checks impede performance
or interfere with algorithms and data structures tuned to be
performant.

\paragraph{Consider all sciences and scales}  Scientific computing often is discussed in
the same context as high performance computing, leading to a focus on
scientific computing problems that demand the scale provided by HPC systems.
Scientific computing occurs at all complexity levels though: small scale data processing
scripts used to perform statistical calculations on a laptop, simple Python models developed
to explore an idea, and so on, are all also valid scientific applications even if they do not
exhibit deep complexity or scale.  Furthermore, there are complex areas of scientific computing
that are not well represented in the flagship applications on HPC systems.  These include
agent-based models, discrete event simulations, probabilistic models, and so on.  To answer the
demands of a wide set of scientific users we should adopt challenge problems from
areas underrepresented by the areas of scientific computing that overlap with HPC: small science
is science too!

\paragraph{Consider validation along with verification}  The formal reasoning
community largely focuses on the problem of verification: does a program
correctly implement a specification that defines its desired behavior.  In the 
sciences we often care equally about the validation problem: have
we written the correct program for our domain problem?  Challenge problems that
attempt to address validation should draw inspiration from efforts within the 
science and engineering communities already looking at verification and validation
through the lens of their disciplines. The ASME verification and validation guidelines~\cite{asme} 
are a good model of these discipline specific practices. It is likely that validation will be harder to find
mechanized challenge problems to study, but given its importance in the field it is
valuable to study.

\section{Conclusion and final thoughts}

In this position paper,
we advocate for a generalizable set of \emph{correctness} challenge problems for scientific computing applications.
We believe that suites of challenge problems can push the state of the art in scientific computing
correctness forward in much the same way that performance challenges have for high performance computing.
These challenges will to help bridge the gap between two relatively disjoint communities: the
formal verification community and the scientific computing community.
Such a challenge set must span the computer abstraction boundaries
used in scientific computing, incorporating both low-level implementation correctness issues (such as memory and numerical safety) as well as high-level scientific domain correctness criteria.
Our recommendations aim to start a dialogue between formal methods and scientific computing domain experts,
motivating a common set of correctness challenges that will enable more practical and powerful formal verification techniques within the scientific computing community.

%% file: paper.bbl
\begin{thebibliography}{10}
\providecommand{\bibitemdeclare}[2]{}
\providecommand{\surnamestart}{}
\providecommand{\surnameend}{}
\providecommand{\urlprefix}{Available at }
\providecommand{\url}[1]{\texttt{#1}}
\providecommand{\href}[2]{\texttt{#2}}
\providecommand{\urlalt}[2]{\href{#1}{#2}}
\providecommand{\doi}[1]{doi:\urlalt{https://doi.org/#1}{#1}}
\providecommand{\eprint}[1]{arXiv:\urlalt{https://arxiv.org/abs/#1}{#1}}
\providecommand{\bibinfo}[2]{#2}

\bibitemdeclare{misc}{inriaGalleryVerified}
\bibitem{inriaGalleryVerified}
\emph{\bibinfo{title}{{G}allery of verified programs ---
  toccata.gitlabpages.inria.fr}}.
\newblock
  \bibinfo{howpublished}{\url{https://toccata.gitlabpages.inria.fr/toccata/gallery/}}.

\bibitemdeclare{misc}{mathcomp}
\bibitem{mathcomp}
\emph{\bibinfo{title}{{M}athematical {C}omponents}}.
\newblock \bibinfo{howpublished}{\url{https://math-comp.github.io/}}.

\bibitemdeclare{misc}{sat}
\bibitem{sat}
\emph{\bibinfo{title}{{S}{A}{T} {C}ompetitions --- satcompetition.github.io}}.
\newblock \bibinfo{howpublished}{\url{https://satcompetition.github.io/}}.

\bibitemdeclare{misc}{nsv3}
\bibitem{nsv3}
 (\bibinfo{year}{2010}): \emph{\bibinfo{title}{NSV-3: Third International
  Workshop on Numerical Software Verification.}}
\newblock
  \bibinfo{howpublished}{\url{https://www.lix.polytechnique.fr/Labo/Sylvie.Putot/NSV3/}}.

\bibitemdeclare{inproceedings}{shonan}
\bibitem{shonan}
\bibinfo{author}{Baris \surnamestart Aktemur\surnameend},
  \bibinfo{author}{Yukiyoshi \surnamestart Kameyama\surnameend},
  \bibinfo{author}{Oleg \surnamestart Kiselyov\surnameend} \&
  \bibinfo{author}{Chung-chieh \surnamestart Shan\surnameend}
  (\bibinfo{year}{2013}): \emph{\bibinfo{title}{Shonan challenge for generative
  programming: short position paper}}.
\newblock In: {\slshape \bibinfo{booktitle}{Proceedings of the ACM SIGPLAN 2013
  Workshop on Partial Evaluation and Program Manipulation}},
  \bibinfo{series}{PEPM '13}, \bibinfo{publisher}{Association for Computing
  Machinery}, \bibinfo{address}{New York, NY, USA}, p.
  \bibinfo{pages}{147–154}, \doi{10.1145/2426890.2426917}.

\bibitemdeclare{misc}{coqnum}
\bibitem{coqnum}
\bibinfo{author}{S.~\surnamestart Aubry\surnameend},
  \bibinfo{author}{S.~\surnamestart Boldo\surnameend},
  \bibinfo{author}{F.~\surnamestart Clement\surnameend},
  \bibinfo{author}{F.~\surnamestart Fassole\surnameend},
  \bibinfo{author}{L.~\surnamestart Leclerc\surnameend},
  \bibinfo{author}{V.~\surnamestart Martin\surnameend},
  \bibinfo{author}{M.~\surnamestart Mayero\surnameend} \&
  \bibinfo{author}{H.~\surnamestart Mouhcine\surnameend}
  (\bibinfo{year}{2025}): \emph{\bibinfo{title}{{M}icaela {M}ayero /
  {N}umerical {A}nalysis in {R}ocq · {G}it{L}ab}}.
\newblock
  \bibinfo{howpublished}{\url{https://depot.lipn.univ-paris13.fr/mayero/rocq-num-analysis/}}.

\bibitemdeclare{inproceedings}{npb1}
\bibitem{npb1}
\bibinfo{author}{D.~H. \surnamestart Bailey\surnameend},
  \bibinfo{author}{E.~\surnamestart Barszcz\surnameend}, \bibinfo{author}{J.~T.
  \surnamestart Barton\surnameend}, \bibinfo{author}{D.~S. \surnamestart
  Browning\surnameend}, \bibinfo{author}{R.~L. \surnamestart
  Carter\surnameend}, \bibinfo{author}{L.~\surnamestart Dagum\surnameend},
  \bibinfo{author}{R.~A. \surnamestart Fatoohi\surnameend},
  \bibinfo{author}{P.~O. \surnamestart Frederickson\surnameend},
  \bibinfo{author}{T.~A. \surnamestart Lasinski\surnameend},
  \bibinfo{author}{R.~S. \surnamestart Schreiber\surnameend},
  \bibinfo{author}{H.~D. \surnamestart Simon\surnameend},
  \bibinfo{author}{V.~\surnamestart Venkatakrishnan\surnameend} \&
  \bibinfo{author}{S.~K. \surnamestart Weeratunga\surnameend}
  (\bibinfo{year}{1991}): \emph{\bibinfo{title}{The NAS parallel
  benchmarks—summary and preliminary results}}.
\newblock In: {\slshape \bibinfo{booktitle}{Proceedings of the 1991 ACM/IEEE
  Conference on Supercomputing}}, \bibinfo{series}{Supercomputing '91},
  \bibinfo{publisher}{Association for Computing Machinery},
  \bibinfo{address}{New York, NY, USA}, p. \bibinfo{pages}{158–165},
  \doi{10.1145/125826.125925}.

\bibitemdeclare{inbook}{npb2}
\bibitem{npb2}
\bibinfo{author}{David~H. \surnamestart Bailey\surnameend}
  (\bibinfo{year}{2011}): \emph{\bibinfo{title}{NAS Parallel Benchmarks}}, pp.
  \bibinfo{pages}{1254--1259}.
\newblock \bibinfo{publisher}{Springer US}, \bibinfo{address}{Boston, MA},
  \doi{10.1007/978-0-387-09766-4_133}.

\bibitemdeclare{article}{bh}
\bibitem{bh}
\bibinfo{author}{Josh \surnamestart Barnes\surnameend} \& \bibinfo{author}{Piet
  \surnamestart Hut\surnameend} (\bibinfo{year}{1986}): \emph{\bibinfo{title}{A
  hierarchical O(N log N) force-calculation algorithm}}.
\newblock {\slshape \bibinfo{journal}{Nature}}
  \bibinfo{volume}{324}(\bibinfo{number}{6096}), pp. \bibinfo{pages}{446--449},
  \doi{10.1038/324446a0}.

\bibitemdeclare{misc}{nist-juliet}
\bibitem{nist-juliet}
\bibinfo{author}{Paul \surnamestart Black\surnameend} (\bibinfo{year}{2018}):
  \emph{\bibinfo{title}{Juliet 1.3 Test Suite: Changes From 1.2}},
  \doi{10.6028/NIST.TN.1995}.

\bibitemdeclare{techreport}{mantevo}
\bibitem{mantevo}
\bibinfo{author}{Paul~Stewart \surnamestart Crozier\surnameend},
  \bibinfo{author}{Heidi~K. \surnamestart Thornquist\surnameend},
  \bibinfo{author}{Robert~W. \surnamestart Numrich\surnameend},
  \bibinfo{author}{Alan~B. \surnamestart Williams\surnameend},
  \bibinfo{author}{Harold~Carter \surnamestart Edwards\surnameend},
  \bibinfo{author}{Eric~Richard \surnamestart Keiter\surnameend},
  \bibinfo{author}{Mahesh \surnamestart Rajan\surnameend},
  \bibinfo{author}{James~M. \surnamestart Willenbring\surnameend},
  \bibinfo{author}{Douglas~W. \surnamestart Doerfler\surnameend} \&
  \bibinfo{author}{Michael~Allen \surnamestart Heroux\surnameend}
  (\bibinfo{year}{2009}): \emph{\bibinfo{title}{Improving performance via
  mini-applications}}.
\newblock \bibinfo{type}{Report}, \doi{10.2172/993908}.

\bibitemdeclare{inproceedings}{fpbench}
\bibitem{fpbench}
\bibinfo{author}{Nasrine \surnamestart Damouche\surnameend},
  \bibinfo{author}{Matthieu \surnamestart Martel\surnameend},
  \bibinfo{author}{Pavel \surnamestart Panchekha\surnameend},
  \bibinfo{author}{Chen \surnamestart Qiu\surnameend},
  \bibinfo{author}{Alexander \surnamestart Sanchez-Stern\surnameend} \&
  \bibinfo{author}{Zachary \surnamestart Tatlock\surnameend}
  (\bibinfo{year}{2017}): \emph{\bibinfo{title}{Toward a Standard Benchmark
  Format and Suite for Floating-Point Analysis}}.
\newblock In \bibinfo{editor}{Sergiy \surnamestart Bogomolov\surnameend},
  \bibinfo{editor}{Matthieu \surnamestart Martel\surnameend} \&
  \bibinfo{editor}{Pavithra \surnamestart Prabhakar\surnameend}, editors:
  {\slshape \bibinfo{booktitle}{Numerical Software Verification}},
  \bibinfo{publisher}{Springer International Publishing}, pp.
  \bibinfo{pages}{63--77}, \doi{10.1007/978-3-319-54292-8_6}.

\bibitemdeclare{article}{dixit1991spec}
\bibitem{dixit1991spec}
\bibinfo{author}{Kaivalya~M. \surnamestart Dixit\surnameend}
  (\bibinfo{year}{1991}): \emph{\bibinfo{title}{Paper: The SPEC benchmarks}}.
\newblock {\slshape \bibinfo{journal}{Parallel Comput.}}
  \bibinfo{volume}{17}(\bibinfo{number}{10–11}), p.
  \bibinfo{pages}{1195–1209}, \doi{10.1016/S0167-8191(05)80033-X}.

\bibitemdeclare{article}{linpack}
\bibitem{linpack}
\bibinfo{author}{Jack~J. \surnamestart Dongarra\surnameend},
  \bibinfo{author}{Piotr \surnamestart Luszczek\surnameend} \&
  \bibinfo{author}{Antoine \surnamestart Petitet\surnameend}
  (\bibinfo{year}{2003}): \emph{\bibinfo{title}{The LINPACK Benchmark: past,
  present and future}}.
\newblock {\slshape \bibinfo{journal}{Concurrency and Computation: Practice and
  Experience}} \bibinfo{volume}{15}(\bibinfo{number}{9}), pp.
  \bibinfo{pages}{803--820}, \doi{10.1002/cpe.728}.
\newblock \eprint{https://onlinelibrary.wiley.com/doi/pdf/10.1002/cpe.728}.

\bibitemdeclare{book}{nsv3journal}
\bibitem{nsv3journal}
\bibinfo{author}{Georgios \surnamestart Fainekos\surnameend},
  \bibinfo{author}{Eric \surnamestart Goubault\surnameend},
  \bibinfo{author}{Sylvie \surnamestart Putot\surnameend} \&
  \bibinfo{author}{Stefan \surnamestart Ratschan\surnameend}
  (\bibinfo{year}{2011}): \emph{\bibinfo{title}{Mathematics in Computer
  Science: Numerical Software Verification}}.
\newblock \bibinfo{volume}{5(4)}, \bibinfo{publisher}{Springer International
  Publishing}.
\newblock
  \urlprefix\url{https://link.springer.com/journal/11786/volumes-and-issues/5-4}.

\bibitemdeclare{book}{salishan}
\bibitem{salishan}
\bibinfo{author}{John~T. \surnamestart Feo\surnameend} (\bibinfo{year}{1992}):
  \emph{\bibinfo{title}{Comparative Study of Parallel Programming Languages:
  The Salishan Problems}}.
\newblock \bibinfo{publisher}{Elsevier Science Inc.}, \bibinfo{address}{USA},
  \doi{10.1016/C2009-0-09810-9}.

\bibitemdeclare{inproceedings}{gallery}
\bibitem{gallery}
\bibinfo{author}{Jean-Christophe \surnamestart Filli{\^a}tre\surnameend} \&
  \bibinfo{author}{Andrei \surnamestart Paskevich\surnameend}
  (\bibinfo{year}{2013}): \emph{\bibinfo{title}{Why3 --- Where Programs Meet
  Provers}}.
\newblock In \bibinfo{editor}{Matthias \surnamestart Felleisen\surnameend} \&
  \bibinfo{editor}{Philippa \surnamestart Gardner\surnameend}, editors:
  {\slshape \bibinfo{booktitle}{Programming Languages and Systems}},
  \bibinfo{publisher}{Springer Berlin Heidelberg}, pp.
  \bibinfo{pages}{125--128}, \doi{10.1007/978-3-642-37036-6_8}.

\bibitemdeclare{article}{asme}
\bibitem{asme}
\bibinfo{author}{Christopher~J \surnamestart Freitas\surnameend}
  (\bibinfo{year}{2020}): \emph{\bibinfo{title}{Standards and methods for
  verification, validation, and uncertainty assessments in modeling and
  simulation}}.
\newblock {\slshape \bibinfo{journal}{Journal of Verification, Validation and
  Uncertainty Quantification}} \bibinfo{volume}{5}(\bibinfo{number}{2}), p.
  \bibinfo{pages}{021001}, \doi{10.1115/1.4047274}.

\bibitemdeclare{phdthesis}{fmm}
\bibitem{fmm}
\bibinfo{author}{Leslie~Frederick \surnamestart Greengard\surnameend}
  (\bibinfo{year}{1987}): \emph{\bibinfo{title}{The rapid evaluation of
  potential fields in particle systems}}.
\newblock Ph.D. thesis, \bibinfo{address}{USA},
  \doi{10.7551/mitpress/5750.001.0001}.
\newblock \bibinfo{note}{AAI8727216}.

\bibitemdeclare{inbook}{verifythis}
\bibitem{verifythis}
\bibinfo{author}{Marieke \surnamestart Huisman\surnameend},
  \bibinfo{author}{Ra{\'u}l \surnamestart Monti\surnameend},
  \bibinfo{author}{Mattias \surnamestart Ulbrich\surnameend} \&
  \bibinfo{author}{Alexander \surnamestart Weigl\surnameend}
  (\bibinfo{year}{2020}): \emph{\bibinfo{title}{The VerifyThis Collaborative
  Long Term Challenge}}, pp. \bibinfo{pages}{246--260}.
\newblock \bibinfo{publisher}{Springer International Publishing},
  \doi{10.1007/978-3-030-64354-6_10}.

\bibitemdeclare{inproceedings}{liao2017dataracebench}
\bibitem{liao2017dataracebench}
\bibinfo{author}{Chunhua \surnamestart Liao\surnameend},
  \bibinfo{author}{Pei-Hung \surnamestart Lin\surnameend},
  \bibinfo{author}{Joshua \surnamestart Asplund\surnameend},
  \bibinfo{author}{Markus \surnamestart Schordan\surnameend} \&
  \bibinfo{author}{Ian \surnamestart Karlin\surnameend} (\bibinfo{year}{2017}):
  \emph{\bibinfo{title}{DataRaceBench: a benchmark suite for systematic
  evaluation of data race detection tools}}.
\newblock In: {\slshape \bibinfo{booktitle}{Proceedings of the International
  Conference for High Performance Computing, Networking, Storage and
  Analysis}}, pp. \bibinfo{pages}{1--14}, \doi{10.1145/3126908.3126958}.

\bibitemdeclare{techreport}{mcmahon1986livermore}
\bibitem{mcmahon1986livermore}
\bibinfo{author}{Frank~H \surnamestart McMahon\surnameend}
  (\bibinfo{year}{1986}): \emph{\bibinfo{title}{The Livermore Fortran Kernels:
  A computer test of the numerical performance range}}.
\newblock \bibinfo{type}{Technical Report}, \bibinfo{institution}{Lawrence
  Livermore National Lab., CA (USA)}.

\bibitemdeclare{article}{simon2005sat2002}
\bibitem{simon2005sat2002}
\bibinfo{author}{Laurent \surnamestart Simon\surnameend},
  \bibinfo{author}{Daniel \surnamestart Le~Berre\surnameend} \&
  \bibinfo{author}{Edward~A \surnamestart Hirsch\surnameend}
  (\bibinfo{year}{2005}): \emph{\bibinfo{title}{The SAT2002 competition}}.
\newblock {\slshape \bibinfo{journal}{Annals of Mathematics and Artificial
  Intelligence}} \bibinfo{volume}{43}, pp. \bibinfo{pages}{307--342},
  \doi{10.1007/s10472-005-0424-6}.

\end{thebibliography}
